\documentclass[12pt]{article}
\usepackage{amsfonts}
\usepackage{amssymb}
\usepackage{cite}
\def\hybrid{\topmargin -20pt    \oddsidemargin 0pt
        \headheight 0pt \headsep 0pt
        \textwidth 6.25in       % A4 paper
        \textheight 9.5in       % A4 paper
        \marginparwidth .875in
        \parskip 5pt plus 1pt   \jot = 1.5ex}

\hybrid

\newcommand{\cD}{{\cal D}}

\newcommand{\cL}{{\cal L}}

\newcommand{\cN}{{\cal N}}
\newcommand{\cO}{{\cal O}}

\newcommand{\beq}{\begin{equation}}
\newcommand{\eeq}{\end{equation}}
\newcommand{\bi}{\begin{itemize}}
\newcommand{\ei}{\end{itemize}}
\newcommand{\bea}{\begin{eqnarray}}
\newcommand{\eea}{\end{eqnarray}}
\newcommand{\ba}{\begin{array}}
\newcommand{\ea}{\end{array}}
\newcommand{\bt}{\begin{tabular}}
\newcommand{\et}{\end{tabular}}
\newcommand{\bc}{\begin{center}}
\newcommand{\ec}{\end{center}}
\newcommand{\half}{\frac12}

\def\S{\hat S}
\def\N{\hat N}
\def\vev#1{\langle #1 \rangle}
\begin{document}

\begin{titlepage}
\begin{center}

\hfill hep-th/0307198\\
%\hfill FSU-TPI-08/02\\
%\hfill CERN-TH/2002-377
%{\Large \bf Draft}

\vskip 1.5cm 
{\large \bf A Note on effective $N=1$ Super Yang-Mills Theories\\
versus Lattice Results}\footnote{Work supported by: 
DFG -- The German
Science Foundation, European RTN Program HPRN-CT-2000-00148 and
the DAAD -- the German Academic Exchange Service.}

\vskip 1.5cm

{\bf David G.\ Cerde\~no, Anke Knauf and Jan Louis}  \\ 

\vskip 20pt

{\em II. Institut f\"ur Theoretische Physik\\
Universit\"at Hamburg\\ 
Luruper Chaussee 149\\
 D-22761 Hamburg, Germany}\\

\vskip 10pt

{email: {\tt  david.garcia.cerdeno, anke.knauf, jan.louis@desy.de}} \\

\vskip 0.8cm

\end{center}

\vskip 2cm

\begin{center} {\bf ABSTRACT } \end{center}
%\vspace{-2mm}

\noindent 

We compare the glueball mass spectrum of an effective $N=1$
pure super Yang-Mills theory formulated 
in terms of a three-form supermultiplet with the available lattice data.
These confirm the presence of four scalars and two Majorana
fermions but the detailed mass spectrum is difficult 
to reconcile with the effective supersymmetric theory.
By imposing supersymmetry and using two of four bosonic masses 
we get a prediction for the remaining masses
as well as the mixing angles. 
We find that the mass of the three-form
dominates over the contribution of the 
Veneziano-Yankielowicz-Dijkgraaf-Vafa term.
As a byproduct we introduce a Fayet-Iliopoulos
term for the three-form multiplet and show that it 
generates a glueball condensate.
\vfill 

%\today
July 2003

\end{titlepage}

%%%%%%%%%%%%%%%%%%%%%%%%%%%%%%%%%%%%%%%%%%%%%%%%%%%
\section{Introduction}
%
%%%%%%%%%%%%%%%%%%%%%%%%%%%%%%%%%%%%%%%%%%%%%%%%%%%%%%%%%%%%
\setcounter{equation}{0}

Recently there has been considerable progress in understanding
some aspects of  strongly coupled
$\cN=1$ supersymmetric gauge theories in four space-time dimensions
below their confinement scale \cite{DV,CDSW,CSW}.
More precisely, corrections to the Veneziano-Yankielowicz
superpotential \cite{VY} have been proposed and conjectured
to give an exact superpotential $W(S)$ in terms of  the glueball
superfield $S = trW^\alpha W_\alpha$.
Among other things these developments strengthened the belief
that an effective action is the appropriate description of 
confined supersymmetric $\cN=1$ gauge theories.

However, it has been pointed out in ref.\ \cite{FGS}
that the Veneziano-Yankielowicz superpotential only
gives rise to mass-terms of the (complex) gluino condensate 
$\langle \lambda\lambda \rangle$
but that the glueballs $\langle F_{\mu\nu} F^{\mu\nu}\rangle$,
$\langle F_{\mu\nu} \tilde F^{\mu\nu}\rangle$
remain massless.
The reason is that in 
the Veneziano-Yankielowicz approach the 
glueball $\langle F_{\mu\nu} F^{\mu\nu}\rangle$
appears in the auxiliary $F$ component of the chiral superfield $S$
and hence no mass term can arise.
$S$ only contains two physical scalars and therefore cannot be adequate
to describe the dynamics of the four bound states
$\langle \lambda\lambda \rangle, \langle \bar\lambda\bar\lambda \rangle,
\langle F_{\mu\nu} F^{\mu\nu}\rangle,
\langle F_{\mu\nu} \tilde F^{\mu\nu}\rangle$.
%In order to include non-zero glueball masses the 
%effective theory has to be formulated in terms of a three-form multiplet $U$.

However, as stressed
in refs.\ \cite{FGS,BGG} 
$S$  really is a constrained chiral multiplet and should better be
viewed as the field strength $S\sim \bar D^2 U$
of a three-form multiplet $U$. Adopting this point of
view it is possible to add a supersymmetric mass term for $U$ 
and in this way introduce two additional massive bosonic
and fermionic degrees of freedom and 
generate  glueball masses \cite{FGS}.
%Dualizing the three-form multiplet the spectrum
%of the low lying excitations can be described by two 
%(instead of one) massive chiral multiplets.

Independent of these developments lattice simulations 
of supersymmetric pure $SU(2)$ gauge theories have been
improved \cite{DMcoll,quenched,pena}.\footnote{%
The results on $SU(3)$ are in
ref.\ \cite{SU3} while the subject is reviewed in \cite{Review}.}
Most of the lattice computations use Wilson-type lattice actions
where supersymmetry is softly broken by a gluino mass term
and later recovered in the continuum limit.
The spectrum of the low-lying glueball- and gluino-condensates 
has been computed and shown
to contain four scalar fields and two Majorana fermions. Furthermore,
supersymmetric Ward identities have been checked 
indicating that supersymmetry is recovered in the continuum limit.

The purpose of this letter is to compare the lattice results 
of \cite{DMcoll} a little more carefully with the approach 
of \cite{FGS}
and show that in order to reach agreement
strong consistency constraints for both the lattice 
simulations and the low energy effective Lagrangian emerge.
We first briefly review the proposal of \cite{FGS} and then compare it
with the lattice simulations. 
By imposing supersymmetry and fitting two bosonic masses
we get a prediction for the remaining masses
as well as the mixing angles. We find that the mass of the three-form
dominates over the contribution of the 
Veneziano-Yankielowicz-Dijkgraaf-Vafa term.
As a byproduct we also introduce a Fayet-Iliopoulos
term for the three-form multiplet and show that it 
generates a glueball condensate $\vev{F_{\mu\nu}F^{\mu\nu}}$.

%%%%%%%%%%%%%%%%%%%%%%%%%%%%%%%%%%%%%%%%%%%%%%%%%%%%%%%%%%%
\section{The Veneziano-Yankielowicz effective action}
\setcounter{equation}{0}
%Let us start by briefly recalling the results of \cite{VY,FGS}.
The starting point is a pure $\cN=1$ supersymmetric $SU(N)$ gauge
theory with a vector multiplet $V$ in the adjoint representation of
$SU(N)$. As physical components it includes a vector field $v_\mu$
and a gluino $\lambda_\alpha$ both in the 
 the adjoint representation of $SU(N)$.
The superspace Lagrangian  reads \cite{WB}
\bea\label{Ldef}
\cL &=& - \frac{i}{8\pi} \int d^2\theta\, \tau\, tr W^\alpha W_\alpha\ 
+h.c.\\
&=& -\frac1{2g^2}\, tr F_{\mu\nu} F^{\mu\nu} 
- \frac{\Theta}{16\pi^2}\, tr F_{\mu\nu} \tilde F^{\mu\nu} 
+\frac{i}{g^2}\, tr \bar \lambda \bar\sigma^\mu D_\mu \lambda\ ,\nonumber
\eea
where $\tau \equiv \frac{\Theta}{2\pi} + i \frac{4\pi}{g^2}$
is the complex gauge coupling and $W_\alpha$ is the
superfield which contains the field strength $F_{\mu\nu}$. It is 
defined in terms of a real vector superfield $V=V^\dagger$ as
\beq\label{Wdef}
W_\alpha = -\frac14 \bar D^2 e^{-V} D_\alpha e^V \ ,
\eeq
Due to its definition it obeys \cite{WB,BGG}
\beq\label{dW}
\bar \cD_{\dot\alpha} W_\alpha = 0 \ , \qquad
\cD^\alpha  W_\alpha = \bar\cD_{\dot\alpha}  \bar W^{\dot\alpha}\ ,
\eeq
where $\cD_\alpha,\cD_{\dot\alpha}$ are
the gauge covariant superspace derivatives.
%\footnote{In terms of
%ordinary superspace derivatives these relations read
%$D_{\dot\alpha} W_\alpha = 0,  

It is believed that this asymptotically free gauge theory confines
below the scale 
\beq
\Lambda\ =\ M\, e^{\frac{2\pi i \tau}{3N}}\ ,
\eeq
where $M$ is some high energy scale at which the gauge theory
(and the coupling $\tau$)
are defined. 
Below the confinement scale $\Lambda$  colorless bound states form
such as the gluino condensates
$\langle\lambda^\alpha\lambda_\alpha\rangle,
\langle\bar\lambda_{\dot\alpha}\bar\lambda^{\dot\alpha}\rangle$ and the 
CP-even and CP-odd glueballs
$\langle F_{\mu\nu} F^{\mu\nu}\rangle, 
\langle F_{\mu\nu} \tilde F^{\mu\nu}\rangle$.
They do not break supersymmetry \cite{index} but they do break
the chiral symmetry $\lambda \to e^{i\kappa}\lambda$
of the original theory (\ref{Ldef}).

Veneziano and Yankielowicz \cite{VY} proposed  an effective description
below the confinement scale $\Lambda$ in terms of a chiral superfield
$S:=tr W^\alpha W_\alpha$ 
with the effective Lagrangian\footnote{%
In \cite{VY,FGS} a different definition is used:
$S= \frac{\beta(g)}{2g}\, tr W^\alpha W_\alpha$ 
where $\beta(g)$ is the (exact) $\beta$-function.
Here we prefer to define $S$ without factors of the gauge coupling
in order to keep the holomorphic properties transparent.
In string theory $\tau$ is not a constant but rather a dynamical chiral superfield.}
\beq
\cL_{\rm eff} =  \int d^4\theta\, K_{\rm eff}(S,\bar S) +
\int d^2\theta\, W_{\rm eff}(S)\ ,
\eeq
where
\beq\label{KWVY}
W_{\rm eff}(S)\ =\ \N \, S\,\big(\ln \frac{S}{\Lambda^3}-1\big)\ ,
\quad
\N\ \equiv\ \frac{N}{32\pi^2}\ .
\eeq
The superpotential $W$ is designed to reproduce the
chiral anomaly. 
$K$ was originally fixed by dimensional
analysis and superconformal anomalies to be
$K_{\rm eff}(S,\bar S) = {\frac1\alpha} (S\bar S)^\frac13$
with $\alpha$ being a dimensionless normalization constant
\cite{VY,Shore}. However, by allowing $K_{\rm eff}$ to explicitly depend 
on $\Lambda$ more general K\"ahler potentials are conceivable.
Therefore in our analysis we will not use a specific $K_{\rm eff}$ but instead express
everything in terms of appropriate derivatives of $K_{\rm eff}$.

In accord with the Witten index \cite{index}
this effective theory has $N$ supersymmetric ground states
determined by $\frac{\partial W_{\rm eff}}{\partial S}=0$
which correspond to\footnote{The $U(1)$ chiral symmetry is not
completely broken by the anomaly but appropriate integer shifts
of $\theta$ leave a discrete ${\bf Z}_{2N}$ intact. 
The gluino condensate is only invariant under
$\lambda\to-\lambda$ and thus it breaks the ${\bf Z}_{2N}$ to ${\bf Z}_2$. 
As a consequence $N$ different ground states appear which are
parameterized by the phase of $S$.}
\beq\label{Svev}
\vev{S}  = \Lambda^3 e^{\frac{2\pi i n}{N}}\ , \quad n = 0, \ldots, N-1\ .
\eeq
The appearance of these $N$ ground states from the minimization of $W$
is a bit tricky and has been discussed in \cite{KS,CSW}.
We return to this issue in our discussion of the three-form multiplet.

Dijkgraaf and Vafa \cite{DV}
added a chiral multiplet in the adjoint representation
of $SU(N)$ to the original pure supersymmetric Yang-Mills theory (\ref{Ldef}).
By giving this multiplet a large mass
it can be integrated out of the effective action but it leaves behind
polynomial corrections in $W$ which are of the form
$W_{\rm eff}(S)\ =\ \N \, S\,\big(\ln \frac{S}{\Lambda^3}-1\big)
+ \N \sum_n a_n S^n $. These corrections shift the location of 
$\vev S$ and also shift the mass term.

$S$ has an expansion in terms of component fields 
\beq\label{SSUN}
S= tr\lambda^\alpha\lambda_\alpha +\ldots 
- \theta^2(tr\frac12  F_{\mu\nu} F^{\mu\nu} + \frac{i}{2}
trF_{\mu\nu} \tilde F^{\mu\nu}+ \ldots) \ ,
\eeq
and thus (\ref{Svev}) implies the formation of the gluino condensate
$\langle\lambda^\alpha\lambda_\alpha\rangle$ while the 
glueball condensate $\langle F_{\mu\nu} F^{\mu\nu}\rangle$ and
$\langle F_{\mu\nu} \tilde F^{\mu\nu}\rangle$ do not form.
Similarly, expanding $W_{\rm eff}$ around $\vev{S}$ one finds a mass
term for $\langle\lambda^\alpha\lambda_\alpha\rangle$ but no
glueball masses.
It is this fact which led Farrar, Gabadaze and Schwetz 
to propose a modification of the VY effective action \cite{FGS}
by formulating the effective theory in terms of a three-form
multiplet $U$.\footnote{An alternative way of generating glueball
  masses was suggested in \cite{SSchechter}.}
The necessity to amend or reformulate the 
VY description had been stressed before in \cite{BGG,KS}.
The common criticism amounts to the fact that the second constraint
in (\ref{dW}) should be taken seriously as a quantum constraint.
In terms of $S=tr W^2$ this constraint reads
\beq
D^2 S - \bar D^2 \bar S = \Omega \ ,
\eeq
where $\Omega$ is a superfield whose lowest component is 
the topological density 
$tr F\tilde F$, i.e.\
\beq
\Omega\ =\ i \epsilon^{\mu\nu\rho\sigma} F_{\mu\nu} F_{\rho\sigma} 
+ \theta \ldots\
=\ - 4i \epsilon^{\mu\nu\rho\sigma} 
\partial_\mu \omega_{\nu\rho\sigma} + \theta \ldots\ ,
\eeq
with $\omega_{\nu\rho\sigma} = tr(v_\nu\partial_\rho v_\sigma - \frac23 i
v_\nu v_\rho v_\sigma)$ being the Chern-Simons three-form.
In other words, the lowest component of $\Omega$ is the field strength
of a three-form. In fact $S$ itself can be viewed as the field strength
supermultiplet of a three-form multiplet $U$.
Before we come to the effective action
let us therefore briefly recall some facts about  the three-form
multiplet  \cite{BGG,Gates}.

\section{The three-form multiplet}
\setcounter{equation}{0}
A real vector superfield $V$ has in its $\theta\bar\theta$-component
a vector field $v_\mu$. However, one can equivalently use the 
Hodge dual three-index antisymmetric tensor or in other words the 
three-form $C_3$ as the $\theta\bar\theta$-component
of a real vector superfield.\footnote{The two fields are related
via $v_\mu \sim \epsilon_{\mu\nu\rho\sigma} C^{\nu\rho\sigma}$.} 
The difference emerges when one considers
the corresponding field strengths which are not dual to each other.
The field strength of a vector superfield $V$ is
the chiral superfield $W_\alpha$ introduced in (\ref{Wdef}) which 
contains $F_{\mu\nu}$ as the $\theta$ component and 
is invariant under the gauge transformations
$V\to V + \Phi + \bar \Phi$ where $\Phi$ is a chiral superfield.
In components this transformation contains the standard gauge transformation
$v_\mu\to v_\mu + \partial_\mu \alpha$. 

Let $U=U^\dagger$ be the superfield which contains $C_3$ 
in its $\theta\bar\theta$-component. Its field strength $S$
is defined by\footnote{%
A generic chiral superfield
$\Phi$ can always be written in terms of an unconstrained 
superfield $X$ as $\Phi = \bar D^2 X,\, \bar \Phi = D^2 \bar X$
but in general $X$ is not real.}
\beq\label{cchiral}
S = -4\bar D^2 U \ , \qquad \bar S = -4 D^2 U\ ,
\eeq
which  is a constrained chiral superfield in that it satisfies
\beq\label{ccon}
\bar D_{\dot\alpha} S = 0 \ , \qquad 
D^2 S -\bar D^2 \bar S = \Omega\ .
\eeq
Here $\Omega$ is a superfield which contains the four-form field strength
$F_4 = dC_3$ in its lowest component.
$S$ (and $\Omega$) are left invariant by the superfield gauge transformation
$U \to U+ L$, where $L$ is a linear multiplet
obeying $D^2 L = \bar D^2 L =0$. At the component level
this corresponds to 
the three-form gauge invariance $C_3\to C_3 + d \Theta_2$ which leaves 
$F_4$ invariant. Note that (\ref{ccon})
does allow the possibility of a supersymmetric VEV $\vev{S}$ since
(\ref{ccon}) is invariant under the 
(supersymmetric) shift $S\to S+ {\rm const.}$.
Thus a more precise version of 
(\ref{cchiral}) reads $S = \vev{S} -4 \bar D^2 U$. We will come back
to this issue in the next section.

A massless three-form contains no physical degree of freedom
since its field strength $F_4$ is dual to a constant.
At the level of superfields this duality states that 
a three-form multiplet is dual to a chiral multiplet with
the physical degrees of freedom being 
a Weyl-fermion and a complex scalar \cite{BGG}.
Since the gauge invariance is unbroken 
the most general action can be expressed in terms of only
the field strength $S$ and reads
\beq\label{Lstand}
\cL = \int d^4\theta K(S,\bar S) + \int d^2\theta\, W(S) + h.c.\ .
\eeq
Thus we see that standard interactions of 
the chiral field strength $S$ describe a massless three-form multiplet.
However, the presence of the three-form does change the 
minimum energy condition. After carefully dualizing the
three-form one finds that the potential is minimized by \cite{BGG}
\beq\label{Wmod}
\frac{\partial \hat W(S)}{\partial S}=0\ ,\qquad
\textrm{where}\qquad  \hat W(S) = W(S) + icS \ . 
\eeq
At the tree level $c$ is a real constant, the dual of $F_4$.
Adding a term $icS$ to $W$ has also been advocated in refs.\ \cite{KS,CSW}
by a different reasoning. Here we see that it naturally appears if one
takes $S$ to be the field strength of a three-form multiplet.
Furthermore, the effective theory is known to have domain wall
solutions interpolating between the $N$ different ground states
of (\ref{Svev}) \cite{DS}. The three-form $C_3$ is the gauge field
which naturally couples to these domain walls. 
The charge satisfies a Dirac-type quantization condition 
which in turn results in the quantization 
$c=\frac{n}{16\pi}, n \in {\bf N}$ \cite{BP,DS}.
Using (\ref{Wmod}) one now finds the $N$ different ground states
displayed in (\ref{Svev}) as the minimum energy condition.

So far we considered a massless three-form multiplet. Let us now
discuss the modifications which appear when a  mass term 
and a Fayet-Iliopoulos term for $C_3$ are included. 
A massive three-form has one physical
degree of freedom which it gains by `eating' an appropriate
Goldstone boson. This Goldstone boson is a two-form $B_2$ with an
invariant coupling $\cL \sim (C_3 - dB_2)^2$. $B_2$ can be removed 
from the Lagrangian by an appropriate gauge transformation.
In a supersymmetric theory $B_2$ resides in  a linear multiplet $G$
and thus  the Lagrangian (\ref{Lstand}) receives the additional terms
\beq\label{mU}
\delta\cL_{m_U} =  - \int d^4\theta\, 
\Big(\frac12 m^2_U (U-G)^2 - \xi (U-G)\Big)\ .
\eeq
We see that the gauge invariance
$U \to U+ L$ can be maintained by assigning the transformation law
$G\to G + L$ to the linear multiplet $G$. Keeping the 
three-form gauge
invariance is crucial since for the $SU(N)$ gauge theory
it is related to ordinary gauge invariance. This can be seen from the fact
that in this case the three-form is nothing but the Chern-Simons
three-form which does transform under the (non-Abelian) gauge symmetry.
In this way the three-form gauge invariance is linked to 
$SU(N)$ invariance.
If one fixes a gauge  $U=U^\prime+G$ (`unitary gauge') $G$ disappears
from the action or in other words it is `eaten' by $U$.
In this gauge the `longitudinal' degrees of freedom of $U$
which are a gauge redundancy in the massless case become physical
degrees of freedom. One bosonic degree of freedom is 
represented by the massive  three-form 
which is dual to a scalar. Supersymmetry requires
that this scalar comes accompanied with an additional
bosonic and two fermionic degrees of freedom originally residing in $G$.
Thus, the massive three-form multiplet has 
altogether four bosonic and four fermionic degrees of freedom,
i.e.\ twice the number of 
of physical degrees of freedom of $S$.\footnote{In \cite{FGS}
it is suggested that the massive three-form multiplet
can be  described equivalently
by two chiral multiplets.}
To see this more explicitly let us now turn to the effective action
suggested in \cite{FGS}.

\section{The effective action of a massive three-form }
\setcounter{equation}{0}

Refs.\ \cite{FGS} propose a modification of the VY effective action
such that also glueball masses can be accommodated. Here we further generalize
this action by also adding a Fayet-Iliopoulos term which will lead
to the possibility of describing a glueball condensate.
The basic idea is to take the constraints (\ref{dW})
seriously also at the quantum level and view $S$ not as a chiral 
field but as a constrained chiral multiplet or in other words
as the field strength of a three-form multiplet.
In this case the basic variable of the effective theory
is not $S$ but rather $U$ and the effective action
should be formulated in terms of $U$. 
Adding ($S$-dependent)
mass- and Fayet-Iliopoulos terms one has\footnote{%
Obviously, one can add higher powers of $U-G$ to $\cL$.
%$\sum_n U^n f_n(S\bar S)$. 
However, such  terms  do not influence the mass spectrum
but correspond instead to additional interactions.}
\beq\label{LFGS1}
\cL = \int d^4\theta \, \Big( K_{\rm eff}(S,\bar S)  
-\frac12 \,
 m^2_U(S,\bar S)\, (U-G)^2 + \xi(S,\bar S)\, (U-G)\Big)
%-  \frac{U^2}{(S\bar S)^\frac13} 
+ \int d^2\theta\, W_{\rm eff}(S)\ +h.c.\ ,
\eeq
where $W_{\rm eff}(S)$ is as in eq.\ (\ref{KWVY}) but 
should now be viewed as a function of $U$.
$K_{\rm eff},m^2_U, \xi$ are arbitrary functions of $S$;
in order to determine the minimum and the mass spectrum of the theory we
do not really need to know their full analytic structure but 
only the first term in a Taylor expansion around $\vev S$.
As we have already stated above, in order to accommodate a 
supersymmetric VEV for $S$ the relation (\ref{cchiral})
has to be modified to $S = \vev{S} - 4\bar D^2 U$. 
Strictly speaking we should use this form of $S$ to derive 
the effective action in components and then determine $\vev S$
by the minimum energy condition. 
As expected this procedure leads again to (\ref{Svev}). 
In order to not overload the notation and
to make the following formulas look more canonical let us chose
the specific vacuum $\vev{S} = \Lambda^3$ right from the beginning and define
\beq\label{Shatdef}
S = \Lambda^3 + \Lambda^2  \S \ , \qquad \S \equiv -4\bar D^2 U  \ , 
\qquad \vev{\S} =0\ .
\eeq
Let us stress that chosing one of the other vacua of (\ref{Svev})
yields an entirely equivalent result. In fact also for the massive three-form
we recover the result of \cite{BGG} that the minimum energy condition
can be expressed as in (\ref{Wmod}).
Using dimensional analysis we can constrain the leading
terms in the Taylor expansion around $\vev S$ of the couplings to be 
\bea\label{Kexp}
K_{\rm eff}(S,\bar S)  &=& k \S\bar{ \S} + \cO((\S\bar{ \S})^2) \ ,\nonumber \\
 m^2_U(S,\bar S) &=& m^2_U+ \cO(\S\bar{ \S})\ , \\ 
\xi(S,\bar S)&=& \xi \Lambda^2 + \cO(\S\bar{ \S})\ ,\nonumber 
\eea
where by slight abuse of 
notation $k, m^2_U, \xi$ now denote constants. 

The next task is to compute the Lagrangian (\ref{LFGS1}) in components
and determine the vacuum and the mass matrices. 
To large extent this was already
done in refs.\ \cite{FGS} and thus we can be very brief in the following.
The only difference compared to \cite{FGS} is that we do not
use a specific $S$-dependence for the couplings $K_{\rm eff}$ and $m^2_U$
since they do not enter the mass matrices. Furthermore, we 
add a Fayet-Iliopoulos term in order to allow for the possibility
of a non-trivial $\vev U$. Following \cite{FGS}
we expand $U$  in component fields as follows
\bea\label{Ucomp}
U&=& \vev B + B + i\theta\chi - i \bar\theta\bar\chi
+\frac1{16}\theta^2 \bar A
+\frac1{16}\bar\theta^2 A
+\frac1{48}\,
\theta\sigma^\mu\bar\theta \epsilon_{\mu\nu\rho\kappa}C^{\nu\rho\kappa} \\
&&+\, \frac1{2}\theta^2\bar\theta\big(
\frac{\sqrt 2}{8} \bar\psi+\bar\sigma^\mu\partial_\mu\chi\big)
+\frac1{2}\bar\theta^2\theta\big(
\frac{\sqrt 2}{8} \psi-\sigma^\mu\partial_\mu\bar\chi\big)
+\frac14\theta^2\bar\theta^2\big(\frac14\Sigma - \partial^\mu\partial_\mu B)\ ,
\nonumber\eea
where 
$B$ is a real and $A$ a complex scalar field, 
$C^{\nu\rho\kappa}$ is the three-form,
$\chi,\psi$ are Weyl fermions and
$\Sigma$ is an auxiliary field.
We have already anticipated the fact 
that the lowest component of $U$ will receive a VEV  due to
the presence of  the Fayet-Iliopoulos term and therefore included 
a term $\vev B$.
Using (\ref{Shatdef}) one identifies the components of $\S$ to be
\beq\label{Scomp}
\S = A + \sqrt{2}\theta \psi + \theta^2 (\Sigma + i F_4) \ ,
\eeq
where 
$F_4 = \frac{1}{3!}\epsilon_{\mu\nu\rho\sigma}\partial^\mu C^{\nu\rho\sigma}$.
In order to have canonically normalized kinetic terms for all fields
we need to further rescale $\S \to k^{-\half} \S$ and 
$B\to \frac{\sqrt 2}{m_U}\, B,\, \chi\to \frac{\sqrt 2}{m_U}\, \chi$. 
Inserted into (\ref{LFGS1}) using (\ref{Kexp}), (\ref{Ucomp}), (\ref{Scomp})
one arrives at a Lagrangian written in terms
of the massive three-form $C_3$ and its field strength $F_4$.
The auxiliary field $\Sigma$ is eliminated by its equation of motion
\beq\label{eom}
\Sigma\ = \
\frac{m_U}{16\sqrt{2k}}\, B - \frac{\N\Lambda}{k}\, {\rm Re} A\ .
\eeq
Finally, dualizing $C_3$ to a scalar $\sigma$ via
\beq
C_{\mu\nu\rho}\ =\ -\frac{16}{m_U} \,\sqrt{\frac{k}{2}}\,
\epsilon_{\mu\nu\rho\kappa}\, \partial^\kappa\sigma \ ,
%\ -\frac{16^2 k}{m_U^2} \,
%\epsilon_{\mu\nu\rho\kappa}\partial^\kappa 
%\Big(8 \sigma + \frac{\N\Lambda}{k}\, {\rm Im}A\Big)\ ,
\eeq
we arrive at
\bea\label{finalL}
\cal{L} & = & - \partial_\mu \Phi^i \partial^\mu \Phi^i
- i \bar{\Psi}^i\bar{\sigma}^\mu\partial_\mu \Psi^i 
-  m^2_{ij} \Phi^i\bar\Phi^j 
-(\frac12\, m_{ij} \Psi^i\Psi^j \ + h.c.)\nonumber\\
&& +\ {\rm higher\ order\ interactions}\ ,
\eea
where
$\phi^i \equiv (A,\frac{1}{\sqrt 2} (B + i\sigma)),\,
\Psi^i \equiv (\psi,\chi),\, i=1,2$
and 
\beq\label{masses}
m_{ij} = \left(\begin{array}{cc}
    m_{11}&m_{12}\\  
    m_{12}&0
\end{array}\right)\ , \qquad m_{11} = \frac{\N\Lambda}{k}\ , \quad 
m_{12}= \frac{m_U}{16\sqrt{k}} \ .
\eeq
(Of course this is exactly the same result obtained in \cite{FGS}
which can be explicitly seen by using the correspondence
$k=\frac{1}{9\alpha} ,\, m_U = \sqrt{2/\delta}\, \Lambda,\, \N = \gamma$.)
Furthermore, a proper minimum of the potential requires
\beq
\vev B =  \xi\, \frac{\Lambda^2}{m_U^2} \ .
\eeq
As anticipated the Fayet-Iliopoulos term $\xi$ induces a VEV
for $B$. In terms of the original $SU(N)$ gauge theory 
$B$ is a mixture of $trF^2$ and $\lambda\lambda$ as can be seen from
(\ref{SSUN}), (\ref{Scomp}), (\ref{eom}).

Let us now turn to a discussion of the mass spectrum.
{}From (\ref{masses}) we see that for the fermion $\psi$ sitting
in $\S$ a Majorana mass term arises directly from the 
VY-superpotential. $m_U$ on the other hand induces a Dirac-mass
term for $\psi$-$\chi$ while no Majorana mass term arises
for $\chi$. In the next section we need 
the eigenvalues of the fermion mass matrix $m_{ij}$ and the bosonic
mass matrix $m_{ij}^2$ which are given by 
\beq\label{eigenvals}
m_f = \frac12 m_{11} \pm \sqrt{m_{12}^2 + \frac14 m_{11}^2}\ \ ,\qquad
m^2_b = \frac12 m_{11}^2 + m_{12}^2 
\pm m_{11} \sqrt{m_{12}^2 + \frac14 m_{11}^2}\ \ .
\eeq

If we consider the correction to the superpotential $W$
computed in \cite{DV} the VEV for $S$ is shifted and an additional
contribution to $m_{11}$ arises.
For instance, after the inclusion of a purely quadratic correction
%$W_{eff}=\hat N S(\ln\frac{S}{\Lambda^3}-1) +a_2S^2$
$\hat N a_2S^2$ to the superpotential,  and parameterizing the
new vacuum expectation value of $S$ by
$
\langle S\rangle=\Lambda^3+\Lambda^2\delta\ ,
$
$m_{11}$ is shifted according to
\begin{equation}\label{DVshift}
m_{11}\to m_{11}\Lambda\left(\frac1{\Lambda+\delta}+2a_2\Lambda^2\right) \ ,
\end{equation}
while $m_{12}$ remains unchanged.

%%%%%%%%%%%%%%%%%%%%%%%%%%%%%%%%%%%%%%%%%%%%%%%%%%%%%%%%%%%%%%%%%%%%%%%%
\section{Comparison with lattice results}
\setcounter{equation}{0}
Various simulations for pure $SU(2)$
super-Yang-Mills theories have been performed on the lattice \cite{DMcoll}.
The two basic issues arising are on the one hand the recovery
of supersymmetry in the continuum limit and on the other hand
the necessity to include dynamical chiral fermions.
Most of the available lattice results use Wilson-type 
lattice actions with a bare  gluino mass term added which breaks
supersymmetry (and the chiral symmetry) softly. 
The gluino mass is then tuned 
such that supersymmetry is recovered in the continuum limit.
Restoration of supersymmetry is checked by 
computing superconformal Ward identities.
% and by observing that the
%low lying excitations assembles in appropriate supermultiplets. 
The lattice simulations of ref.\ \cite{DMcoll}
show a non-trivial mass spectrum for four scalar degrees of freedom 
and two Majorana fermions  which seem to assemble
in two chiral multiplets near the supersymmetric 
limit. Let us focus on the bosonic states.
The lightest states are the CP-even glueball $B$ 
(called $0^+$ in \cite{DMcoll})
and the CP-odd gluino-condensate ${\rm Im} A$ 
(called $a-\eta^\prime$ in \cite{DMcoll})
and they are almost degenerate in mass.
The  CP-odd glueball $\sigma$ ($0^-$)
and the CP-even gluino-condensate ${\rm Re} A$ ($a-f_0$) 
are also degenerate in mass and heavier than $B$ and ${\rm Im} A$.
Furthermore, the mass difference
between these two sets of states is much larger than the value of the
gluino mass
used in the simulation
and therefore does not appear to be an effect of the softly
broken supersymmetry.

However, taken at face value this is in conflict 
with basic properties of supersymmetry.
{} From eq.\ (\ref{finalL}) we learn that the four bosonic
states combine into two complex scalar fields $A$ and 
$\frac{1}{\sqrt 2} (B + i\sigma)$. In terms of the original
$SU(N)$ gauge theory $A$ corresponds to the gluino condensate
$tr\lambda\lambda$ while $B+ i\sigma$ is a mixture of  
$trF^2 + i tr F\tilde F$ with $A$ (c.f.\ eqs. (\ref{SSUN}), (\ref{Scomp}), (\ref{eom})). 
The two complex scalar fields $A$ and $\frac{1}{\sqrt 2} (B + i\sigma)$
do mix via the mass matrix (\ref{masses}) and therefore 
the mass eigenstates are a linear combination of $A$ and $B+ i\sigma$
with generically different masses (c.f.\ eq.\ (\ref{eigenvals})).
However, the mixing is only among complex scalar fields 
as can be seen from (\ref{finalL}) in that no terms $\Phi^2$ or
$\bar\Phi^2$ appear. Therefore
the CP-even and CP-odd states of the same  complex scalar
continue to be degenerate in mass. It is in fact a fundamental
property of unbroken supersymmetry  that  CP-even and CP-odd states
in the same multiplet have to be degenerate in mass.\footnote{%
This is related to the 
R-symmetry of the super-algebra which preserves the complex structure
and pairs CP-even and CP-odd scalars into a complex scalar field.}
The supermultiplets can mix  but the mass
eigenstates again have to be supersymmetric multiplets
and therefore the CP-even and CP-odd states of the same multiplet
are mass degenerate.
Thus, $A$ can mix with  $B+ i\sigma$ but the resulting
mass eigenstates cannot lead to a mass split between ${\rm Im} A$
and ${\rm Re} A$.
In the this respect the lattice results which show
$B$ and ${\rm Im} A$ mass degenerate
seem to be in conflict with  unbroken supersymmetry.

There appear to be various possible resolutions of this puzzle.
First of all it could be that 
due to the mixing the states in the lattice simulations have
been misidentified. In the lattice simulation
correlation functions of operators with given quantum numbers
are computed for large
Euclidean times where they are dominated by the lightest states
with these quantum numbers. This can be used to extract the mass
of the states. Thus, when computing correlation function
of, say $tr F^2$, it is in principle possible that this
correlation function is dominated by the CP-even gluino-ball
$a-f_0$. This would imply a mixing between the bosonic
states with the same parity which, however, is not 
observed in the lattice simulations \cite{DMcoll}.
Furthermore, experience from QCD appear to make this possibility
unlikely \cite{GMprivate}.

The second possibility is that in the supersymmetric limit 
all four states are really degenerate in mass 
and that the $0^+$ glueball $B$ 
and the CP-odd gluino-condensate $a-\eta^\prime$ (${\rm Im} A$)
are really in different
but almost degenerate supermultiplets. 
Since the more reliable lattice measurements
are the two light states this would mean that
the masses of the heavy states 
have to considerably decrease as one comes closer
to the supersymmetric limit.
However, improved lattice simulations currently under way
do not indicate any tendency in this direction \cite{GMprivate}.

Thus there remains a  puzzle when comparing the mass spectrum
obtained in lattice simulation with computations based
on supersymmetric effective actions. The fact that the conflict
between the two approaches is at such a fundamental level
makes this only the more interesting.
Estimating finite size effects both on the lattice and analytically
might shed some light on this puzzle.

Let us now take the second solution
seriously and ask what we can learn for the  mass spectrum
if we insist on supersymmetry and 
only take the two low lying states into account.
In this case all states have to be almost degenerate
and
we can determine the parameters $k$ and 
$m_U$. This  `predicts' the masses for the $0^-$ and $a-f_0$
as well as their mixing angle. 
{}From eq.\ (\ref{eigenvals}) we learn that a 
degeneracy among all four bosonic and fermionic
states  is not so easy to achieve and 
requires 
\beq 
m_{12} \gg m_{11} \quad \Rightarrow \quad
m_{U} \gg \frac{\N}{\sqrt k}\, \Lambda \ .
\eeq
Thus the mass of the three-form $m_{U}$ has to be the dominant contribution 
in the mass matrices
and is far more important than the contribution arising from the
Veneziano-Yankielowicz-Dijkgraaf-Vafa superpotential. 
This also implies that once the Dijkgraaf-Vafa correction is taken into account $m_{11}$
given in (\ref{DVshift}) cannot be large.
Furthermore,
{}from (\ref{masses}) we see that 
in this limit the mixing angle of the bosons is minimal while
the fermions mix maximally.

%%%%%%%%%%%%%%%%%%%%%%%%%%%%%%%%%%%%%%%%%%%%%%%%%%%%%%%%%

\section{Conclusions}\label{}
\setcounter{equation}{0}
In this letter we expanded on a suggestion put forward in
refs.\ \cite{BGG,FGS} 
to reformulate the effective action
of strongly coupled supersymmetric gauge theories
in terms of a (massive) three-form multiplet in order to account for
glueball masses.
A gauge invariant formulation of the mass term
requires the presence of additional degrees of freedom
and doubles the spectrum compared to the original
Veneziano-Yankielowicz effective theory.
Indeed the lattice simulations do measure
four massive scalars and two Majorana fermions
and in this sense confirm the proposal of \cite{FGS}.
However, taken at face value the observed mass spectrum 
is incompatible with supersymmetry
in that the CP-even and CP-odd part of the complex scalars
show different masses. We discussed two possible resolutions of this puzzle.
Either there is a misidentification of states or all
states have to be almost degenerate in mass.
This in turn requires that the mass term of the three-form
is the dominant contribution and  no mixing among the bosonic states 
occurs while the fermions have to be maximally mixed.
It would be worthwhile to further improve the lattice simulation
and shed light on this puzzle.

We also introduced a Fayet-Iliopoulos term $\xi$ into the theory and showed
that it can lead to a non-trivial glueball condensate 
$\vev{tr F^2}\neq 0$. It would also be nice to measure
$\langle \lambda\lambda \rangle$ and 
$\vev{tr F^2}$ on the lattice and determine in this way
$\Lambda$ and $\xi$.

\vskip 1cm
\clearpage
%%%%%%%%%%%%%%%%%%%%%%%%%%%%%%%%%%%%%%%%%%%%%%%%%%%%%%%%%%%%%%%%
{\large \bf Acknowledgments}

This work is supported by DFG -- The German Science Foundation -- 
within the ``Schwerpunktprogramm Stringtheorie'' and by the
European RTN Program HPRN-CT-2000-00148 and the DAAD -- the German
Academic Exchange Service.

We have benefited from discussions with 
R.~Dijkgraaf, R.~Grimm, V.~Kaplunovsky, R.~Kirchner, L.~Lellouch, A.~Micu,
I.~Montvay, G.~M\"unster, H.-P.~Nilles, C.~Pena, S.~Theisen, C.~Vafa and 
H.~Wittig.

%%%%%%%%%%%%%%%%%%%%%%%%%%%%%%%%%%%

\end{document}